\documentclass[reprint,amsmath,amssymb,notitlepage,prb,longbibliography]{revtex4-2}
\usepackage{graphicx}
\usepackage[whole]{bxcjkjatype} %pdflatexでの日本語用

\usepackage{bm}% bold math
\usepackage{mathrsfs}
\usepackage{multirow}
\usepackage{arydshln}
\usepackage{physics}
\usepackage{comment}
\usepackage{cases}
\usepackage{hyperref}

\usepackage{color}

\allowdisplaybreaks[1]

\newcommand{\average}[1]{\ensuremath{\langle#1\rangle} }

\newcommand{\red}[1]{#1}

\newcommand{\del}[1]{}

\begin{document}

	\title{Spin current generation due to differential rotation}

	\author{Takumi Funato${}^{1,2}$, Shunichiro Kinoshita${}^{3,4}$, Norihiro Tanahashi${}^4$, Shin Nakamura${}^4$, and Mamoru Matsuo${}^{2,5,6,7}$}
	\affiliation{${}^1$Center for Spintronics Research Network, Keio University, Yokohama 223-8522, Japan}
	\affiliation{%
		${}^2$Kavli Institute for Theoretical Sciences, University of Chinese Academy of Sciences, Beijing, 100190, China.
	}%
    \affiliation{${}^3$Department of Physics, College of Humanities and Sciences, Nihon University, Tokyo 156-8550, Japan} 
	\affiliation{${}^4$Department of Physics, Chuo University, Tokyo 112-8551, Japan}
	\affiliation{${}^5$CAS Center for Excellence in Topological Quantum Computation, University of Chinese Academy of Sciences, Beijing 100190, China
	}%
	%\email{}
	\affiliation{${}^6$RIKEN Center for Emergent Matter Science (CEMS), Wako, Saitama 351-0198, Japan}
	\affiliation{
		${}^7$Advanced Science Research Center, Japan Atomic Energy Agency, Tokai, 319-1195, Japan
	}
	
	\date{\today}
	
	\begin{abstract}
We study nonequilibrium spin dynamics in differentially rotating systems, deriving an effective Hamiltonian for conduction electrons in the comoving frame. In contrast to conventional spin current generation mechanisms that require vorticity, our theory describes spins and spin currents arising from differentially rotating systems regardless of vorticity. We demonstrate the generation of spin currents in differentially rotating systems, such as liquid metals with Taylor-Couette flow. 
Our alternative mechanism will be important in the development of nanomechanical spin devices.
	\end{abstract}
	
	\pacs{Valid PACS appear here}
	\maketitle
	
{\it Introduction}. %---
The physics of spin currents,  initially introduced as the flow of Fermi particles~\cite{Breit1938}, 
%[Breit 1938] 
and the diffusion of spin magnetic moments~\cite{Hone1961,Leggett1965}
%[Hone 1961, Leggett 1965], 
and Ising spins~\cite{Kawasaki1966},
has long attracted the interest of researchers.
Generating and controlling spin currents is a key challenge in spintronics~\cite{maekawa2017spin}, involving mechanisms such as %like 
the spin Hall effect~\cite{dyakonov1971Currentinduced,dyakonov1971Possibility,hirsch1999Spin,kato2004Observation,saitoh2006Conversion}, spin pumping~\cite{berger1996Emission,MIZUKAMI20011640,Mizukami_2001,ingvarsson2002Role,lubitz2003Increase}, spin Seebeck effect~\cite{uchida2008Observation}, spin accumulation at ferromagnet/nonmagnet interfaces~\cite{johnson1985Interfacial,valenzuela2006Direct}, and Edelstain effect~\cite{bychkov1984Oscillatory,edelstein1990Spin,sanchez2013Spintochargea}. Advances in nanofabrication now enable mechanical motions in materials for spin transport~\cite{uchida2011Surfaceacousticwavedriven,weiler2012Spin,deymier2014Phononmagnon,puebla2020Acoustic,almisba2020AcousticWaveInduced,camara2019FieldFree,tateno2020Electrical,kawada2021Acoustic,sasaki2017Nonreciprocala,tateno2020Highly,xu2020Nonreciprocal,kuss2021Nonreciprocal,matsumoto2022Large,lyons2023Acoustically,chowdhury2017Nondegeneratea,alekseev2020Magnons,liao2023ValleySelective}. The gyromagnetic effect~\cite{barnett1915Magnetization,a.einstein1915,scott1962review,wallis2006AppliedPhysicsLetters,zolfagharkhani2008NatureNanotecha,harii2019NatCommun,mori2020Phys.Rev.B,chudo2014Observation,chudo2015Rotational,arabgol2019Phys.Rev.Lett.,chudo2021Barnett,wood2017magnetic,hirohata2018magneto,dornes2019Nature,ganzhorn2016NatCommunb,adamczyk2017global}, involving angular momentum interconversion between mechanical rotation and spin, is crucial as it enables spin current generation without strong spin-orbit coupling~\cite{takahashi2016Spin,kobayashi2017Spin,takahashi2020Giant,tabaeikazerooni2020Electron,tabaeikazerooni2021Electrical,tateno2021Einstein}.

The gyromagnetic effect, initially understood through the conservation of angular momentum~\cite{barnett1915Magnetization,a.einstein1915,barnett1935Gyromagnetic,scott1962review}, is now recognized as originating from spin-rotation coupling, $H_\mathrm{sr} = -\boldsymbol{s} \cdot \boldsymbol{\Omega}$, where $\boldsymbol{s}$ is the spin and $\boldsymbol{\Omega}$ is the angular velocity of rigid rotation, derived from the Dirac equation~\cite{hehl1990inertial}. This coupling is analogous to the Zeeman effect, with angular velocity acting as an effective magnetic field on the spin. Theoretical and experimental work shows that  
inhomogeneities of the angular velocity can generate a spin current via the Stern-Gerlach type effect~\cite{matsuo2013Mechanical,matsuo2017Theory,matsuo2017SpinMechatronics}, using vorticity gradients in liquid metal~\cite{takahashi2016Spin,takahashi2020Giant,tabaeikazerooni2020Electron,tabaeikazerooni2021Electrical} and surface acoustic waves~\cite{kobayashi2017Spin,kurimune2020Highly,kurimune2020Observation,tateno2020Electrical,tateno2021Einstein}.  
In these studies, a spin-vorticity coupling, $H_{\rm sv} = - \boldsymbol{s}\cdot \boldsymbol{\omega}$, was explored by replacing constant angular velocity $\boldsymbol{\Omega}$ in $H_{\rm sr}$ with vorticity $\boldsymbol{\omega} = (1/2)\boldsymbol\nabla \times \boldsymbol{v}$. 
However, in addition to the spin-vorticity coupling,
differential rotation $\boldsymbol{\Omega}(\boldsymbol{r}) = \boldsymbol{r} \times \boldsymbol{v}/r^2$ offers a new method for localizing the interaction. This overlooked coupling may enhance our understanding of spin transport driven by non-uniform rotation.

In this study, we investigate the non-equilibrium spin dynamics in differentially rotating systems within a microscopic theory.
By mapping into a comoving frame, we construct an effective Hamiltonian for conduction electrons in these systems, demonstrating the emergence of effective gauge fields. Furthermore, we derive microscopic expressions for the spin density and spin current of conduction electrons driven by these emergent gauge fields. 
The mechanism of spin current generation proposed in the present Letter is based on the fundamental principles of quantum mechanics, without interjecting phenomenological arguments. 

Although our mechanism applies to general differentially rotating systems, we present specific examples of experiments that may verify our proposal.
By applying our mechanism to a liquid metal and a non-magnetic metallic cantilever as examples of differentially rotating systems, we estimate the concrete amount of the spin current. In particular, we show that even in cases such as Taylor-Couette flow where the vorticity-gradient is zero, spin currents can be generated due to the differential rotation.  
Consequently, we uncover mechanisms of angular momentum transfer that have not been captured by
traditional frameworks.

	{\it Emergent gauge fields in comoving frame.} We consider the free electron system subject to momentum scattering and spin-orbit scattering due to the impurities.
    In the inertial laboratory frame the Hamiltonian is given by 
    \begin{align}
	    \hat H' &= \int d^3x \hat \psi^{\dagger}(\bm x) \biggl\{
	     - \frac{\hbar^2}{2m}\bm\nabla^2 - \epsilon_F +V'_{\text{imp}}(\bm x,t) \nonumber \\
	    &\quad
     + \lambda_{\text{so}}\bm \sigma \cdot [ \bm\nabla V'_{\text{imp}}(\bm x,t) \times (-i\hbar\bm\nabla) ]
	    \biggr\}
	    \hat \psi(\bm x),
	    \label{H_lab}
	\end{align}
    where $\hat \psi(\bm x)$ is the electron field operator, $\epsilon_F$ is the Fermi energy, $\bm \sigma =(\sigma^x,\sigma^y,\sigma^z)$ are the Pauli matrices, and $\lambda_{\text{so}}$ is the strength of the spin-orbit interaction.
    The third term represents the impurity scattering and the fourth term represents the spin-orbit scattering.
	Here, $V'_{\text{imp}}(\bm x,t) = \sum_j u[\bm x-\bm r'_j(t)]$ is the total impurity potential, where $u[\bm x-\bm r'_j(t)]$ is a single impurity potential due to the $j$th impurity located at the position $\bm r'_j(t)$.
    It is worth noting that the electrons are subject to the moving impurities because we suppose the total system is differentially rotating.
    To characterize the differential rotation of the system, we introduce a rotation angle $\Phi(\bm{x},t)$ around the $z$ axis, which is chosen as a rotation axis.
    When we take a cylindrical coordinate system, the coordinate transformation from the laboratory frame 
    $\bm r' = (r', \varphi', z')$ to the rotating frame $\bm r=(r,\varphi,z)$ can be written as 
    $r=r'$, $z=z'$, and $\varphi=\varphi'-\Phi(\bm{r}',t)$.
    Note that $\Phi$ is independent of $\varphi$, i.e., $\partial_\varphi \Phi = 0$ because of axisymmetry.
    Supposing $\Phi(\bm{x},t)=0$ at an initial time $t=0$, the position of the $j$th impurity at $t$ 
    is given by $\bm r'_j(t) = \mathcal{R}_z[\Phi(\bm{r}_j,t)] \bm{r}_j$, where $\mathcal{R}_z$ denotes rotation around the $z$ axis and $\bm{r}_j$ is the position at 
   $t=0$.

    Now, we define a generator of the differential rotation with angle $\Phi(\bm{x},t)$ as 
    \begin{equation}
        \hat{Q}_\Phi (t) = \int d^3x \Phi(\bm x, t)
        \hat \psi^{\dagger}(\bm x) J^z \hat \psi(\bm x),
    \end{equation}
    where $J^z$ is the total angular momentum operator acting on coordinates and spin space as 
    $J^z=-i\hbar\partial_\varphi + \hbar\sigma^z/2$.
    Note that $J^z$ and $\Phi(\bm x,t)$ are commutative.
    For an arbitrary state vector in the laboratory frame $|\Psi'(t)\rangle$, the state vector in the rotating frame is given by 
    \begin{equation}
        |\Psi(t)\rangle = \exp \left[\frac{i}{\hbar}\hat{Q}_\Phi(t)\right] |\Psi'(t)\rangle .
    \end{equation}
    The Schr\"odinger equation in the laboratory frame, 
    $i\hbar\partial_t |\Psi'(t)\rangle = \hat H'|\Psi'(t)\rangle$, yields 
    \begin{equation}
        \begin{aligned}
        i\hbar \frac{\partial}{\partial t} |\Psi(t)\rangle &= 
        (e^{i\hat{Q}_\Phi/\hbar} \hat H' e^{-i\hat{Q}_\Phi/\hbar} - \hat{Q}_{\partial_t\Phi}) |\Psi(t)\rangle \\
        &= \hat{H}_T |\Psi(t)\rangle ,
        \end{aligned}
    \end{equation}
    where $\hat{Q}_{\partial_t\Phi} = \int d^3x \partial_t\Phi(\bm x, t) \hat \psi^{\dagger}(\bm x) J^z \hat \psi(\bm x)$ and 
    $\hat{Q}_\Phi$ commute because of $\partial_\varphi\Phi = 0$.
    The Hamiltonian $H_T$ governs dynamics in the rotating frame.
    The density operator in the rotating frame, $\hat \rho(t)$, is given by $\hat \rho(t)=e^{i\hat Q_\Phi/\hbar} \hat \rho'(t) e^{-i \hat{Q}_\Phi/\hbar}$, 
    where $\hat \rho'(t)$ is the density operator in the laboratory frame.
    The time evolution of $\hat{\rho}(t)$ is determined by $i\hbar \partial_t \hat \rho(t) = [\hat H_T, \hat \rho(t)]$.
    Assuming that  
    the single impurity potential $u(\bm{x})$ is isotropic and its typical range $a$, such that $u(\bm{x}) \simeq 0$ for $|\bm{x}|\gg a$, is 
    much smaller than a typical scale of the gradient of the differential rotation, i.e., $a |{\bm \nabla} \Phi| \ll 1$, 
    the Hamiltonian in the rotating frame can be rewritten as 
    \begin{gather}
	\!\!\!\!\!\!    \hat H_T =\!\! \int \!\!d^3x \hat \psi^{\dagger}(\bm x) \biggl\{
	       \frac{(-i\hbar {\bm \nabla}\!\! -\!\! {\bm A}_{s}J^z)^2  }{2m}
% \frac{1}{2m}(-i\hbar {\bm \nabla} - {\bm A}_{s}J^z)^2 
       -\!\! A_{s,0} J^z \!\!- \epsilon_F 
         +\!V_{\text{imp}}(\bm x) \nonumber \\
	    + \lambda_{\text{so}}\bm \sigma \cdot 
     [ {\bm \nabla} V_{\text{imp}}(\bm x) \times (-i\hbar {\bm \nabla} - {\bm A}_{s}J^z ) ]
	    \biggr\}
	    \hat \psi(\bm x),
	    \label{H_rot}
	\end{gather}
    where the time and spatial derivatives of the rotation angle are denoted by  
    \begin{align}
	    A_{s,\mu}(\bm x,t) = \Bigl( \partial_t \Phi(\bm x,t), {\bm \nabla} \Phi(\bm x,t)  \Bigr)
     \quad (\mu = 0,x,y,z).
	    \label{eq:gauge}
	\end{align}
We call $A_{s,\mu}(\bm x,t)$ ``emergent gauge field'' in this Letter. 
    In the rotating frame, the effects of the differential rotation are represented by the emergent gauge fields, 
    whereas the impurity potential given by 
    $V_{\text{imp}}(\bm x) = \sum_j u(\bm x-\bm r_j)$ does not depend on time under the assumption $a |{\bm \nabla} \Phi| \ll 1$.

    {\it Setup}. We present the Fourier representation of the total Hamiltonian in the rotating frame to facilitate calculations: 
    $\hat H_T = \hat H_0 + \hat H_{\text{imp}} + \hat H_{\text{so}} + \hat H'(t)$, 
    where $\hat H'(t)$ is the contribution of the emergent gauge field, and we treat it as a perturbation.
    The first term $\hat H_0 = \sum_{\bm k} \epsilon_{\bm k} \hat \psi_{\bm k}^{\dagger} \hat \psi_{\bm k}$ represents the kinetic term, where $\epsilon_{\bm k}=\hbar^2k^2/2m-\epsilon_F$ is the kinetic energy, and $\hat \psi_{\bm k}$ is the Fourier component of the electron annihilation operator.
	The second and third terms describe the momentum scattering and the spin-orbit scattering due to the impurities, respectively.
	These are expressed as $\hat H_{\text{imp}} = \sum_{\bm k \bm k'} V_{\bm k-\bm k'} \hat \psi_{\bm k}^{\dagger} \hat \psi_{\bm k'}$ and $\hat H_{\text{so}} = i\hbar \lambda_{\text{so} }\sum_{\bm k\bm k'} V_{\bm k-\bm k'} (\bm k\times \bm k') \cdot \hat \psi_{\bm k}^{\dagger} \bm \sigma \hat \psi_{\bm k'}$, 
    where $V_{\bm k}$ denotes the Fourier component of the impurity potential $V_{\text{imp}}(\bm x)$. 
	We assume a short-range impurity potential, i.e., $u(\bm x-\bm r_j)=u_{\text i} \delta(\bm x-\bm r_j)$, where $u_{\text i}$ is the strength of the impurity potential defined by $u_{\text i} = \int d^3x u(\bm x)$ in general, 
    while the perturbed part, denoted by $\hat H'(t)=\hat H_s+\order{L^z}$ with $L^z = -i\hbar \partial_{\varphi}$ being the orbital angular momentum, represents the effect of the emergent gauge fields. 
    The Hamiltonian $\hat{H}_s$ incorporates the electron spin, given by
    \begin{align}
        \hat H_s &= - \frac{\hbar^2}{2m}
        \sum_{\bm k\bm k'\bm q} 
        \hat \psi_{\bm k_+}^{\dagger}
        \left(
        \bm k \sigma^z \delta_{\bm k\bm k'} - \frac{1}{2} \bm A_{s,\bm k-\bm k'} 
        \right)
         \hat \psi_{\bm k'_-}
        \cdot \bm A_{s}(\bm q)
        \nonumber \\
         &\quad
         -\frac{\hbar}{2} \hat s(\bm q) A_{s,0}(\bm q)
        ,
        \label{eq:hs}
    \end{align}
    where $\bm A_{s,\bm q}$ is the Fourier component of the emergent gauge fields, and $\bm k_{\pm}=\bm k\pm \bm q/2$ are defined.
    
    To define the spin-current operator, we consider the temporal modulation of the $z$-polarized spin density, $\partial_t \hat s(\bm q) = - i\bm q \cdot \hat{\bm j}_s(\bm q) + \hat{\mathcal T}_{\bm q}$, where $\hat s(\bm q)=\sum_{\bm k}\hat \psi^{\dagger}_{\bm k_-} \sigma^z \hat \psi_{\bm k_+}$ is the spin-density operator, and $\hat{\mathcal T}_{\bm q}$ describes the spin torque due to the spin-orbit interaction of the impurities.
    The spin-current density operator polarized in the $z$ direction is defined by $\hat{\bm j}_s(\bm q) = \sum_{\bm k\bm k'}\hat \psi^{\dagger}_{\bm k_-'} \bm j_{s,\bm k'\bm k} \hat \psi_{\bm k_+}$, where the matrix elements $\bm j_{s,\bm k'\bm k}$ are given by
    \begin{align}
        \!\!\!\!\!\bm j_{s,\bm k'\bm k}
    	    = \delta_{\bm k'\bm k} \bm v_{\bm k} \sigma^z \!\!+\! \lambda_{\text{so}} V_{\bm k' -\bm k} [\bm e_z 
         \!\!
         \times \!\!(\bm k'\!\!-\bm k)]\! -\!\frac{\hbar \bm A_{s,\bm k'-\bm k}}{2m} ,
    \end{align}
    where $\bm v_{\bm k}=\hbar \bm k/m$ is the velocity and $\bm e_z$ is the unit vector in $z$ direction.

    {\it Calculation of spin current.} We now compute the spin current induced by the emergent gauge fields.
    The statistical averages of the spin density and spin current are given by
    \begin{align}
        \!\!
        \average{\hat{ j}_{\mu}(\bm q,\omega)} =
        \!\!
        \int^{\infty}_{-\infty} 
        \!\!        
        \frac{d\epsilon}{2\pi i} \sum_{\bm k\bm k'} \Tr \left[ j_{s\mu,\bm k'\bm k} G^<_{\bm k_+,\bm k'_-}(\epsilon_+,\epsilon_-) \right],
    \end{align}
    where $\epsilon_{\pm} = \epsilon \pm \omega/2$, $j_{s0,\bm k'\bm k} = \sigma^z \delta_{\bm k'\bm k}$, and the trace is taken for the spin space.
    Here, the four-vector $\hat j_{\mu}=(\hat s,\hat{\bm j}_s)$ represents the spin density and spin current operators.
    The function $G^<_{\bm k_+,\bm k'_-}(\epsilon_+,\epsilon_-)$ is the lesser component of the nonequilibrium path-ordered Green's function, defined by 
    $G_{\bm k,\bm k'}(t,t') = -i\average{T_K \hat \psi_{\bm k_+}(t) \hat \psi^{\dagger}_{\bm k'_-}(t')}$, 
    where $T_K$ is a path-ordering operator, $\hat \psi(t)=\hat U^{\dagger}(t)\hat \psi(t)\hat U(t)$ is the Heisenberg representation with $\hat U(t) = T\text{exp}[-(i/\hbar)\int^{t}_{-\infty} \hat H_T(\tau) d\tau]$ and $T$ being time-ordering operator, and $\average{\cdots}=\text{tr}(\hat \rho \cdots)$  
    is the expectation value with the density operator $\hat \rho$.
    
    Assuming that the characteristic energy scales of the momentum scattering and the spin-orbit scattering due to the impurities are much smaller than the Fermi energy, i.e., $n_{\text i}u_{\text i}\ll \epsilon_F$ and 
    $\hbar^2 \lambda_{\text{so}}^2k_F^4\ll 1$, we treat them in the Born approximation. 
    With the uniformly random distribution of impurities, we perform the average of their positions to obtain the retarded/advanced Green's function: 
    $g^{r/a}_{\bm k}(\epsilon) = 1/(\epsilon  -\epsilon_{\bm k}\pm i\hbar \gamma)$, 
	where $\hbar \gamma=\pi n_{\text i}u_{\text i}^2\nu_0(1+2\hbar^2\lambda_{\text{so}}^2k_F^4/3)$ is the damping constant calculated with the density of state per spin at Fermi level $\nu_0=mk_F/2\pi^2\hbar^2$. 
We assume that $\hbar \gamma \ll \epsilon_F$. This condition is well-satisfied when $u_{\text i} \nu_0 \lesssim 1$.

	The spin-current density in linear response to the emergent gauge fields is expressed as
    $\average{\hat j_\mu(\bm q,\omega)} = K_{\mu\nu}(\omega) A_{s,\nu}(\bm q,\omega)$, where $K_{\mu\nu}(\omega)$ is the response function.
    It is presumed that the time and spatial variation of the differential rotation are much slower than the electron mean-free path $l=v_F\tau$ and momentum relaxation time $\tau=1/2\gamma$, respectively, i.e., 
    $l|\bm \nabla \Phi|\ll1$ and $\tau |\partial_t \Phi| \ll1$, 
    where $v_F=\hbar k_F/m$ is the Fermi velocity and $k_F=\sqrt{2m\epsilon_F/\hbar^2}$ is the Fermi wavenumber.
    In terms of Fourier space, conditions $lq \ll 1$ and $\tau \omega \ll 1$ hold.
    By including the ladder vertex corrections due to the impurities, and using the relations $\hbar v_F q/2 \ll \hbar \gamma \ll \epsilon_F$ and $\hbar \omega/2 \ll \hbar \gamma \ll \epsilon_F$, the response function is calculated as
    \begin{align}
        &K_{\mu \nu}(\omega) = \delta_{\mu 0} \delta_{\nu 0} \frac{\hbar \sigma_0}{2e^2D} \nonumber \\ 
        &\qquad +  i\omega \frac{ \hbar}{4\pi} \sum_{\bm k} v_{\bm k,\mu} \Tr[\sigma^z g^r_+ \sigma^z(v_{\bm k,\nu}+\Lambda_{\nu}^{s})g^a_-],
    \label{response}    
    \end{align}
    where $\sigma_0 = n_ee^2 \tau /m$ is the Drude conductivity with $n_e = 4\epsilon_F \nu_0 /3$ being the number density of the electrons and $e(>0)$ being the elementary charge, and $D=v_F^2\tau /3$ is the diffusion constant.
    We set $v_{\bm k,0} =1$ and $v_{\bm k,i} = \hbar k_i /m$.
Here, $\Lambda^s_{\nu}$ describes the three-point vertex corrections, and $g^{r/a}_{\pm} = g^{r/a}_{\bm k_\pm}(\pm \omega/2)$ are specified.
The first term of the response function represents the spin susceptibility for the rigid rotation~\cite{funato2018Generation}, known as the Barnett effect.

Performing a straightforward calculation, we derive the rotation-induced spin density and spin current:
\begin{align}
    \average{\hat{s}(\bm q,\omega)} &= -i\omega \frac{\hbar \sigma_0}{2e^2D} \frac{\tau_s^{-1}}{Dq^2 - i\omega + \tau_s^{-1}} \Phi (\bm q,\omega),
    \label{eq:spin_density}
    \\
    \average{\hat{\bm j}_s(\bm q,\omega)} &= i\omega \frac{\hbar \sigma_0}{2e^2} \frac{\tau_{s}^{-1}}{Dq^2-i\omega + \tau_{s}^{-1}} i{\bm q} \Phi (\bm q,\omega),
    \label{eq:spin_current}
\end{align}
where $\tau_{s} = 9\tau/8 \hbar^2 \lambda_{\text{so}}^2k_F^4$ is the spin-relaxation time. 
Combining these results, 
%in the real space, 
we obtain Fick's law, 
${\bm j}_s=-D \bm \nabla s$.
This implies that our spin current is a diffusive flow produced by the gradient of the spin density, in which the impurity scattering governs the diffusion.

Now, we focus on long-term dynamics such that time scales are longer than the period of the rotation, $\omega \lesssim \Omega$.
In metals, the spin relaxation time and the spin diffusion length 
are much shorter than the typical scale of the period and that of the spatial variation for the differential rotation, respectively
(i.e., $\Omega^{-1}\gg \tau_s$ and $|\bm\nabla \Omega /\Omega|^{-1} \gg l_s$, where $l_s = \sqrt{D\tau_s}$). 
Thus, the relations $Dq^2,\omega \ll \tau_{s}^{-1}$ are satisfied, and 
the rotation-induced spin current reduces to the following form in the real space:
        \begin{align}
            \bm j_s(\bm x,t) = -\frac{\hbar \sigma_0}{2e^2} \bm \nabla \partial_t \Phi (\bm x,t).
            \label{js}
        \end{align}
In addition, the rotation-induced spin density reduces to  
\begin{equation}
    s(\bm x,t) = \frac{\hbar \sigma_0}{2e^{2}D } \partial_t \Phi (\bm x,t) ,
    \label{s}
\end{equation}
which is the Barnett effect generalized to differential rotations.
The susceptibility given by $\hbar \sigma_0/2e^{2}D $ is identical to that of the Barnett effect for rigid rotations.
These results suggest that the spin density and current are polarized along the rotation axis and the spin current 
is driven in the direction of the spatial gradient of the angular velocity.
By contrast, if the spin relaxation is so slow that $\tau_{s}^{-1} \ll \omega \lesssim \Omega$, the spin density (\ref{eq:spin_density}) as well as the spin current (\ref{eq:spin_current}) vanish,
which implies that the spin relaxation is necessary to generate the spin current and spin density. 
Despite this fact, the magnitude of the spin current (\ref{js}) is 
independent of the spin relaxation time.

The absence of $\tau_{s}$ from the long-term dynamics of the spin density and the spin current is explained as follows.
In the response function (\ref{response}), 
the first term that originates from the spin-rotation coupling $A_{s,0}J^{z}$ in (\ref{H_rot}) is principal, while the other terms including the spin-orbit coupling are suppressed by $\tau_{s}\omega \ll 1$.
This means that the spin density is determined only by the susceptibility of the Barnett effect and the angular velocity. 
The gradient of this spin density produces the spin current due to the diffusion caused by the impurity scattering, as shown.
Thus, the spin density and current are independent of $\tau_s$.

The spin-orbit interaction contributes only to the transient process that is necessary to drive the system to the final steady state, but it does not contribute to long-term dynamics. 
Indeed, for $\omega \gtrsim \Omega$, (\ref{eq:spin_density}) and (\ref{eq:spin_current}) provide the following spin transport equation: 
\begin{equation}
    \frac{\partial s}{\partial t}+{\bm \nabla} \cdot {\bm j}_s=-\frac{s}{\tau_{s}} + \frac{\hbar \sigma_0}{2e^{2}D \tau_s}\partial_t\Phi ,
\label{eq:spin_cons_viloation}
\end{equation}
which describes the transient process with a time scale $\omega \simeq \tau_s^{-1}$.
We expect to obtain similar diffusive spin currents as long as 
there is  
an interaction producing a transient process satisfying $\Omega \ll \tau_s^{-1} \ll \tau^{-1}$, not necessarily that presented here.

\begin{figure}
        \centering
        \includegraphics[width=75mm]{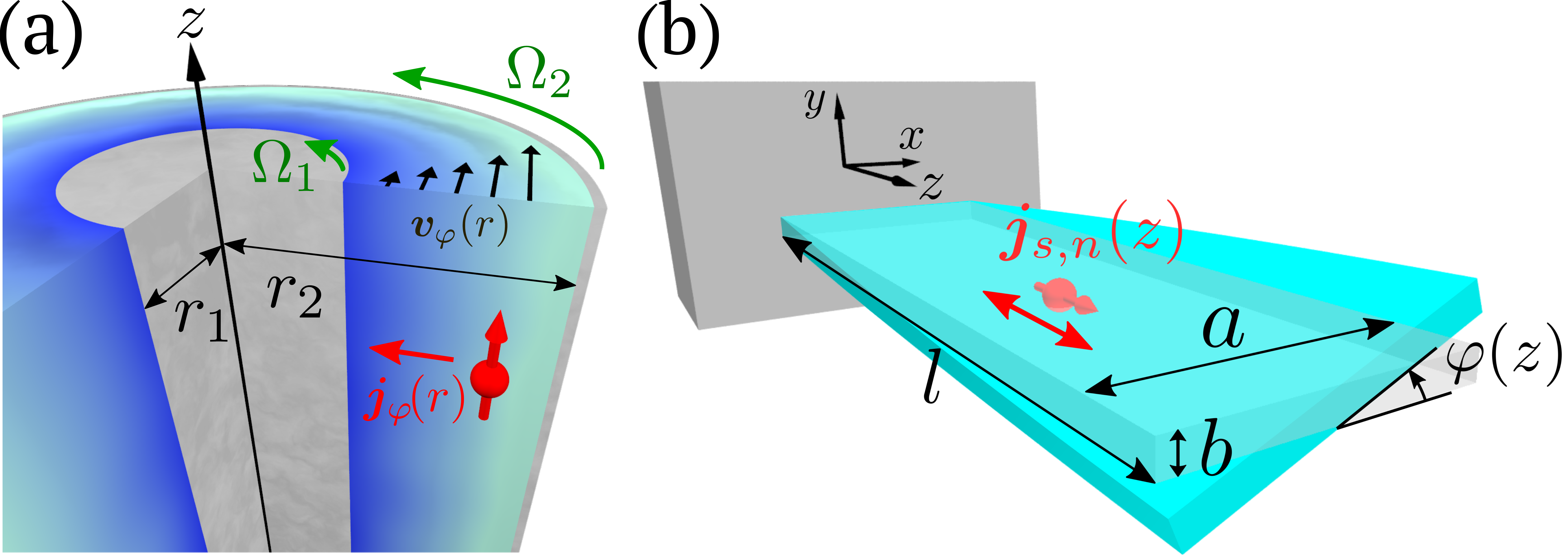}
        \caption{Schematic illustration showing the generation of spin current due to (a) the Taylor-Couette flow in a liquid metal and (b) torsional motion of a cantilever.
        }
        \label{fig:proposal}
    \end{figure}

    {\it Taylor-Couette flow.} As an explicit example, let us consider a two-dimensional steady flow with concentric circular streamlines. 
    In this case, the flow velocity is parallel to the $\varphi$-direction, $\bm v=(0,v_{\varphi},0)$, satisfying the following Navier-Stokes equation: $\partial_r^2 v_{\varphi} + (\partial_r v_{\varphi})/r - v_{\varphi}/r^2=0$. 
    (The detailed derivation is given in Supplemental Material.) %~\cite{supp}
    The general solution is $v_{\varphi} = c_1/r + c_2r$ with integration constants $c_1$ and $c_2$ determined by boundary conditions. 
    The first term represents irrotational flow, while the second term represents rigid-rotation flow. 
    We consider the two infinitely long coaxial cylinders of radii $r_1$ and $r_2$ ($r_2>r_1 >0$), and the inner and outer cylinders are rotating at constant angular velocities $\Omega_1$ and $\Omega_2$, respectively.
    Under these boundary conditions,
    $v_{\varphi}(r_1)=r_1\Omega_1$ and $v_{\varphi}(r_2)=r_2\Omega_2$,
    the constants are obtained as 
     $c_1= (\Omega_1 - \Omega_2) r_1^2r_2^2/(r_2^2-r_1^2)$ and $c_2 = (\Omega_2 r_2^2 - \Omega_1 r_1^2)/(r_2^2-r_1^2)$. 
    This concentric steady flow, known as the Taylor-Couette flow~\cite{taylor1923VIII}, induces the steady differential rotation with angular velocity $\Omega (r) = c_1/r^2 + c_2$, leading to the generation of spin current [see Fig.~\ref{fig:proposal}(a)]:
    \begin{align}
       \bm j_s(r) = \bm e_r \frac{\hbar \sigma_0}{e^2} \frac{ r_1^2r_2^2}{r_2^2-r_1^2}\frac{\Omega_1 - \Omega_2}{r^3},
    \label{TCspincurrent}   
    \end{align}
    where $\bm e_r$ is the unit vector in the $r$ direction. 
Notably, since the vorticity in this system is constant, $\bm \nabla \times \bm v=2c_2\bm e_z$, the conventional spin currents owing to the spin-vorticity coupling, which require the vorticity gradient \cite{matsuo2017Theory,takahashi2016Spin} or time-dependent vorticity \cite{matsuo2013Mechanical}, do not appear. 
On the other hand, our theory predicts the generation of spin current even in vorticity-free cases $c_2 = 0$.

    To estimate the magnitude of the spin current, we assume that the radii of the two cylinders are much larger than the gap between them $d=r_2-r_1$, i.e., $r_1,r_2\gg d$,
    and only the outer cylinder is rotating, $\Omega_1 = 0$ and $\Omega_2 \neq 0$, for simplicity.
    Under this assumption, the spin current is approximated as $j_{s}=\hbar \sigma_0\Omega_2/2e^{2}d$.
    We consider (Ga,In)Sn as the fluid with the electric conductivity $\sigma_0=3.26\times 10^6(\Omega \text m)^{-1}$~\cite{tabaeikazerooni2020Electron}.
    Set $d\sim 1\mu$m and $\Omega_2 \sim 10^2$kHz, the magnitude of the spin current in charge current units is estimated as
   $ej_{s} \sim 1.07 \times 10^2 \mathrm{A} \mathrm{m}^{-2}$.

     {\it 
    Torsional oscillation of cantilever.} As another example, we focus on the torsional oscillation of a cantilever~\cite{wallis2006AppliedPhysicsLetters,harii2019NatCommun}, wherein one of the ends is securely fixed while external forces are exerted on the opposite end. 
    These forces induce only a twisting motion in the cantilever, not bending or other deformations.
    In this case, the angular velocity of the system varies along the rotation axis rather than the radial direction.
    This position-dependent rotation induces relative motion of the impurities in the cantilever, and our analysis applies even to such a setting.
    
    The distortion angle $\varphi(z,t)$ of the cantilever dictates the subsequent equation of motion: $C\partial_z^2 \varphi = \rho_m I \partial_t^2 \varphi$, 
    where $C$ is the torsional rigidity, $\rho_m$ is the mass density, and $I$ is the moment of inertia of the cross section about its center of mass.
    By solving the equation of motion under the boundary conditions $\varphi(0,t)=0$ and $\partial_z \varphi(l,t)=0$ and considering the initial conditions $\varphi (l,0) = \varphi_0$ and $\partial_t \varphi (z,0)=0$, we derive the solution as $\varphi_n(z,t) = \varphi_0 \sin k_nz \cos \omega_n t$, where $k_n=(2n-1)\pi/2l$ and $\omega_n = vk_n$ with the integer $n\geq 1$ and the velocity $v=\sqrt{C/\rho_m I}$.
    Plugging $\varphi$ into $\Phi$ in~(\ref{js}), we find
    the spin current, driven by the $n$th torsional oscillation of cantilever, flows along the $z$ direction as given by [see Fig.~\ref{fig:proposal}(b)]
    \begin{align}
        \bm j_{s,n} (z,t) = \bm e_z\frac{\hbar \sigma_0 \varphi_0v}{2e^2}  k_n^2 \cos k_nz \sin \omega_nt.
    \end{align}
    The mechanism under investigation in this study represents a universal phenomenon, irrespective of material choice, and fundamentally distinct from the previous theory~\cite{fujimoto2020Magnon} that focus solely on magnetic materials.
    
    Finally, we estimate the magnitude of the spin current driven by the torsional oscillation.
    For a plate-shaped cantilever with width $a$, thickness $b$ and length $l$, the quantities $C$ and $I$ are calculated as $C\simeq \mu ab^3/3$ and $I\simeq a^3b/12$ ($a\gg b$) with Lam\'e constant $\mu$.
    The magnitude of the total spin current in charge current units is denoted by $J_{n}=eabj_{s,n}(0,0)$, while that attributed to the first torsional oscillation mode is given by $J_1=(\pi^2 \hbar \sigma_0 \varphi_0/4e)(b/l)^2 \sqrt{\mu/\rho_m}$.
    We consider that the cantilever is composed of copper with weak spin-orbit interaction.
    By using the charge conductivity $\sigma_0=6.45 \times 10^7 \, \Omega^{-1}\text m^{-1}$, the Lam\'e constant $\mu = 48.3 \, \text{GHz}$ and the mass density $\rho_m = 8.96 \times 10^3\, \text{kg}/\text m^3$, the total spin current is estimated as $J_1 \sim 0.15 \, \mu$A for $\varphi_0\sim 0.01$ and $b/l =1/4$.
    
{\it Conclusion and discussion.} We have proposed a mechanism for spin current generation 
in differentially rotating systems. This mechanism produces spin currents even in vorticity-free systems and differs from the known mechanisms based on the spin-vorticity coupling. 
Our result is not phenomenological, but is derived from a microscopic theory of non-equilibrium spin dynamics. This is a natural extension of the Barnett effect in rigidly rotating systems to non-uniform rotating systems. It removes the previous limitation of spin current generation, which was restricted to systems with vorticity, and broadens the potential for spin devices using a wider range of spatially non-uniform mechanical motions. In addition, it would provide a deeper understanding of the ubiquitous gyromagnetic effect, both from a theoretical and an experimental point of view.

In more detail, 
distinctions between our mechanism and those proposed in other literature~\cite{matsuo2013Mechanical,matsuo2017Theory,takahashi2016Spin} can be 
found in
the source terms that violate the conservation of the spin density, 
where the spin current is proportional to the gradient of those source terms. In our case,  
the spin transport equation is given in (\ref{eq:spin_cons_viloation}) and the source term is proportional to $\Omega$, the angular velocity of the orbital rotation around a fixed axis.
On the other hand, the source term given in Ref.~\cite{matsuo2013Mechanical} and those in Refs.~\cite{matsuo2017Theory,takahashi2016Spin} are proportional to $\partial_{t}\tilde{\omega}$ and 
$\zeta\tilde{\omega}$, respectively, where $\tilde{\omega}$ is the vorticity and $\zeta$ is a phenomenological parameter.  
The difference between our proposal and that of Ref.~\cite{matsuo2013Mechanical} 
can be detected experimentally, for the torsional oscillation, from the frequency dependence of the spin current.
We can distinguish our mechanism from those of Refs.~\cite{matsuo2017Theory,takahashi2016Spin} experimentally based on its dependence on phenomenological parameters (for details, see Supplemental Material). %\cite{supp}
Moreover, an irrotational flow, in which the vorticity is zero, can prevent the spin current generation due to the spin-vorticity coupling. 
Creating an irrotational flow akin to a differentially rotating system allows us to detect spin currents specific to our mechanism. One may achieve such flow in the Taylor-Couette flow by taking appropriate boundary conditions that realize $c_{2}=0$ with $c_{1}\neq 0$.
Furthermore, we can distinguish the mechanisms even in the case of $c_{2}\neq 0$. 
Since the vorticity in this system is uniform and time-independent, neither the source term in Ref.~\cite{matsuo2013Mechanical} nor that in Refs.~\cite{matsuo2017Theory,takahashi2016Spin} can contribute to the spin current.
As a result, the Taylor-Couette flow
can generate the spin current in our mechanism but not in the other ones.
Exploring the implications of our mechanism in various experiments would be valuable 
for both fundamental physics and next-generation devices, such as integrating liquid metal fluids with micro-electromechanical systems (MEMS)~\cite{khoshmanesh2017liquid} to utilize electrical and spin degrees of freedom.

%\begin{acknowledgments} 	
	{\it Acknowledgments.} We would like to thank D. Oue and Y. Nozaki for the valuable and informative discussion.
	This work was partially supported by JST CREST Grant No.~JPMJCR19J4, Japan.
	We acknowledge JSPS KAKENHI for Grants (No.~JP21H01800, No.~JP21H04565, No.~JP23H01839, No.~JP21H05186, No.~JP19K03659, No.~JP19H05821, No.~JP18K03623, and No.~JP21H05189). 
The work was supported in part by
the Chuo University Personal Research Grant.
The authors thank RIKEN iTHEMS NEW
working group for providing the genesis of this collaboration.
%\end{acknowledgments} 

\bibliography{differential_rotation.bib}

\clearpage

\begin{widetext}

	\section{Transformation to differentially rotating frame}
    By using the total angular momentum operator acting on coordinates and spin space, %$J^z$,
    \begin{equation}
        J^z \equiv -i\hbar(\bm{x}\times\boldsymbol{\nabla})_z \otimes I
        + 1 \otimes \frac{\hbar}{2}\sigma^z ,
        ,
    \end{equation}
    we define the following operator: 
    \begin{equation}
        \hat{Q}_\Phi (t) = \int d^3x \Phi(\bm x, t)
        \hat \psi^{\dagger}(\bm x) J^z \hat \psi(\bm x),
    \end{equation}
    which is a generator of the differential rotation with angle $\Phi(\bm{x},t)$ around the $z$-axis.
    Note that we assume the rotation angle is axisymmetric, i.e., $\partial\Phi/\partial\varphi = 0$.
    The canonical commutation relation $\{\hat\psi(\bm x), \hat\psi^{\dagger}(\bm y)\}=\delta(\bm x - \bm y)$ 
    yields 
    \begin{equation}
        [\hat{Q}_\Phi (t), \hat\psi(\bm x)] = - \Phi(\bm x, t) J^z \hat\psi(\bm x) ,\quad
        [\hat{Q}_\Phi (t), \hat\psi^{\dagger}(\bm x)] = \Phi(\bm x, t) (J^z \hat\psi(\bm x))^{\dagger} .
    \end{equation}
    We have 
    \begin{equation}
    \begin{aligned}
        \exp\left[\frac{i}{\hbar}\hat{Q}_\Phi(t)\right] \hat\psi(\bm x)
        \exp\left[-\frac{i}{\hbar}\hat{Q}_\Phi(t)\right]
        &= e^{-i\Phi(\bm{x},t)J^z/\hbar} \hat\psi (\bm x) \\
        &= e^{-i\Phi(\bm{x},t)\sigma^z/2} \hat\psi (\bm x')  
        \qquad (\bm{x}' \equiv e^{-\Phi(\bm{x},t) (\bm{x}\times\boldsymbol{\nabla})_z}\bm{x}) .
    \end{aligned}   
    \end{equation}
    Note that it can be explicitly written in matrix form as 
    \begin{equation}
        e^{-\Phi(\bm{x},t) (\bm{x}\times\boldsymbol{\nabla})_z}\bm{x} = \mathcal{R}^{-1}_z[\Phi(\bm{x},t)] \bm{x} , \quad
        \mathcal{R}_z(\varphi) \equiv \mqty( \cos\varphi & -\sin\varphi & 0 \cr \sin\varphi & \cos\varphi & 0 \cr 0 & 0 & 1 ) .
    \end{equation}
    The free part of the Hamiltonian (namely, the kinetic term) transforms as 
    \begin{equation}
        \begin{aligned}
            e^{i\hat{Q}_\Phi/\hbar} \hat{H}_0 e^{-i\hat{Q}_\Phi/\hbar}
            &= \hat{H}_0 + \frac{i}{\hbar}\int^1_0 d\lambda 
            e^{i\lambda\hat{Q}_\Phi/\hbar} [\hat{Q}_\Phi, \hat{H}_0]
            e^{-i\lambda\hat{Q}_\Phi/\hbar} \\
            &= \hat{H}_0 + \frac{i}{\hbar} [\hat{Q}_\Phi, \hat{H}_0]
            - \frac{1}{2\hbar^2} [\hat{Q}_\Phi, [\hat{Q}_\Phi, \hat{H}_0]]
            - \frac{i}{3!\hbar^3} [\hat{Q}_\Phi, [\hat{Q}_\Phi, [\hat{Q}_\Phi, \hat{H}_0]]] + \cdots \\
            &= \frac{1}{2m}\int d^3\boldsymbol{x} 
            [- i\hbar \boldsymbol{\nabla} \hat{\psi}(\boldsymbol{x}) 
            - \boldsymbol{\nabla} \Phi (\boldsymbol{x},t) J_z \hat{\psi}(\boldsymbol{x})
            ]^\dagger \cdot
            [- i\hbar \boldsymbol{\nabla} \hat{\psi}(\boldsymbol{x}) 
            - \boldsymbol{\nabla} \Phi (\boldsymbol{x},t) J_z \hat{\psi}(\boldsymbol{x})
            ] ,
        \end{aligned}
    \end{equation}
    where we have used the following equations: 
    \begin{equation}
        [\hat{Q}_\Phi, \hat{H}_0] = \frac{i\hbar^2}{2m}
        \int d^3\boldsymbol{x}
        \boldsymbol{\nabla} \Phi (\boldsymbol{x},t) \cdot 
        \left[
            ( - i \boldsymbol{\nabla} \hat{\psi}(\boldsymbol{x}))^\dagger
            J_z \hat{\psi}(\boldsymbol{x}) 
            + (J_z \hat{\psi}(\boldsymbol{x}))^\dagger
            ( - i \boldsymbol{\nabla} \hat{\psi}(\boldsymbol{x}))
        \right] ,
    \end{equation}
    \begin{equation}
        [\hat{Q}_\Phi , [\hat{Q}_\Phi , \hat{H}_0]] = - \frac{\hbar^2}{m}
        \int d^3\boldsymbol{x}
        |\boldsymbol{\nabla} \Phi (\boldsymbol{x},t)|^2 
        (J_z \hat{\psi}(\boldsymbol{x}))^\dagger
        J_z \hat{\psi}(\boldsymbol{x}) ,
    \end{equation}
    \begin{equation}
        [\hat{Q}_\Phi ,[\hat{Q}_\Phi , [\hat{Q}_\Phi , \hat{H}_0]]] = 0 .
    \end{equation}
    The impurity potential part is 
    \begin{equation}
        \begin{aligned}
            e^{i\hat{Q}_\Phi/\hbar} \left[\int d^3x \hat \psi^{\dagger}(\bm x) V'_{\text{imp}}(\bm x,t) \hat \psi(\bm x) \right] e^{-i\hat{Q}_\Phi/\hbar}
            &= \int d^3x \hat \psi^{\dagger}(\bm x') V'_{\text{imp}}(\bm x,t) \hat \psi(\bm x') \\
            &= \sum_j \int d^3x \hat \psi^{\dagger}(\bm x') u(\bm x - \bm r'_j(t)) \hat \psi(\bm x')
            \quad (\bm r'_j(t) = \mathcal{R}_z[\Phi(\bm{r}_j,t)] \bm{r}_j) \\
            &= \sum_j \int d^3x \hat \psi^{\dagger}(\bm x) u(\mathcal{R}_z[\Phi(\bm{x},t)] \bm x - \mathcal{R}_z[\Phi(\bm{r}_j,t)] \bm{r}_j) \hat \psi(\bm x) \\
            &\simeq \sum_j \int d^3x \hat \psi^{\dagger}(\bm x) u(\bm x - \bm r_j) \hat \psi(\bm x)
            = \int d^3x \hat \psi^{\dagger}(\bm x) V_{\text{imp}}(\bm x) \hat \psi(\bm x) ,
        \end{aligned}
    \end{equation}
    where we have changed a variable of integration as $\bm x \to \mathcal{R}_z[\Phi(\bm{x},t)] \bm x$ in the third line.
    Because we assume that a single impurity potential $u(\bm x)$ is isotropic and has compact support in $|\bm x|<a$, we have 
   \begin{equation}
       u(\mathcal{R}_z[\Phi(\bm{x},t)] \bm x - \mathcal{R}_z[\Phi(\bm{r}_j,t)] \bm{r}_j) 
       = u(\bm x - \mathcal{R}_z[\Phi(\bm{r}_j,t) - \Phi(\bm{x},t)] \bm{r}_j) \simeq u(\bm x - \bm{r}_j)
   \end{equation}
   for $|\bm{x} - \bm{r}_j| < a \ll 1/|\boldsymbol{\nabla}\Phi|$.
    In a similar manner, the spin-orbit interaction part is 
   \begin{equation}
        \begin{aligned}
            &e^{i\hat{Q}_\Phi/\hbar}
            \left[
                \lambda_{\text{so}}
                \int d^3x \hat \psi^{\dagger}(\bm x) 
                \bm \sigma \cdot [ \boldsymbol{\nabla} V'_{\text{imp}}(\bm x,t) \times (-i\hbar\boldsymbol{\nabla}) ]
                \hat \psi(\bm x)
            \right]
            e^{-i\hat{Q}_\Phi/\hbar} \\
            & = \lambda_{\text{so}}
            \int d^3x \hat \psi^{\dagger}(\bm x)
	           \bm \sigma \cdot 
             [ \boldsymbol{\nabla} V_{\text{imp}}(\bm x) \times (-i\hbar\boldsymbol{\nabla}- \boldsymbol{\nabla}\Phi(\boldsymbol{x},t) J^z ) ]
	    \hat \psi(\bm x) .
	   \end{aligned}
    \end{equation}

	\section{Definition of spin current operator}
	In this section, we define the spin-current density operator through the continuum equation with respect to the spin density.
	The $z$-polarized spin density operator is defined by
	\begin{align}
	    \hat s (\bm x) \equiv \hat \psi^{\dagger}(\bm x) \sigma^z \hat \psi(\bm x).
	\end{align}
	The time derivative of the $z$-polarized spin density operator in the Heisenberg representation is given by
	\begin{align}
	    \pdv{t}\hat s(\bm x,t)
	    = - \bm\nabla \cdot \hat{\bm j}_s(\bm x,t) + \hat{ \mathcal T}(\bm x,t). 
	\end{align}
	The first term is the divergence of the spin-current density:
	\begin{align}
	    \hat{\bm j}_{s}(\bm x) = \hat \psi^{\dagger}(\bm x) \qty[ \frac{\hbar \sigma^z}{2im}  \overleftrightarrow{\bm\nabla}  + \lambda_{\text{so}} \bm e_z\times \bm\nabla V - \frac{\hbar}{2m} \bm A_s ] \hat \psi (\bm x).
	\end{align}
	The second term represents the spin torque due to the impurity spin-orbit interaction:
	\begin{align}
	    \hat{\mathcal T}(\bm x) = -\frac{i}{\hbar} \lambda_{\text{so}} \hat \psi^{\dagger}(\bm x) \qty[ \sigma^z \bm e_z \times \qty( \bm\nabla V \times \frac{\hbar}{i} \bm\nabla ) ] \hat \psi(\bm x).
	\end{align}

	\section{calculation of spin density and spin current density}

\begin{figure}
    \centering
    \includegraphics[width=100mm]{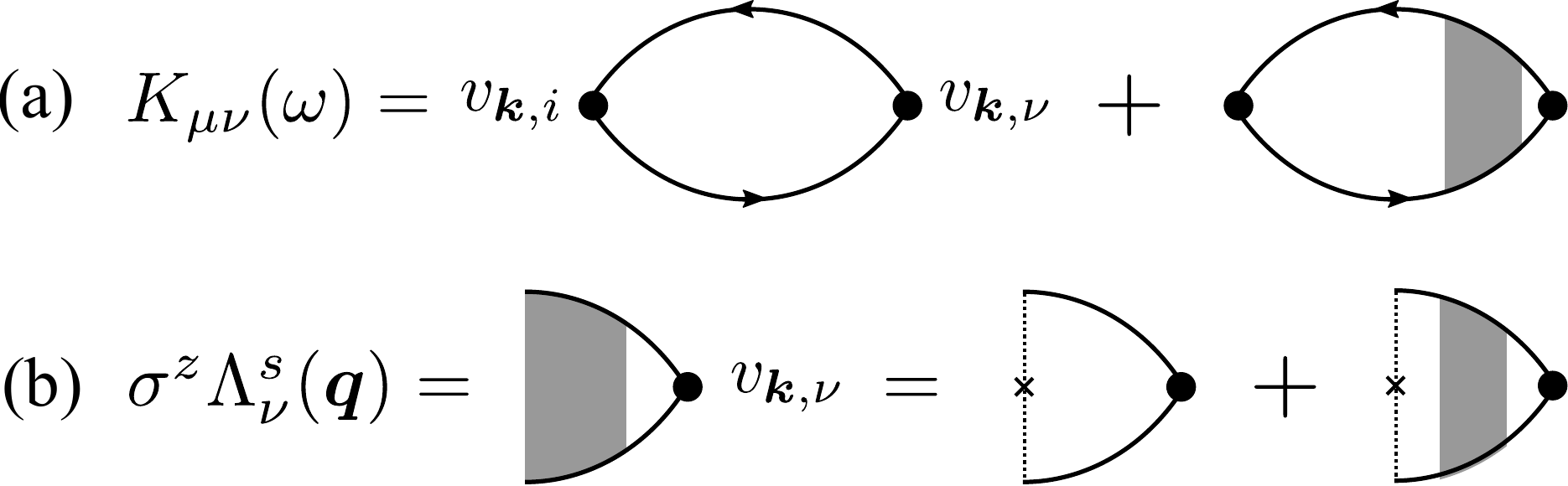}
    \caption{The diagrams representing the (a) response functions and the (b) three-point vertices.
    The black circles represent the vertices and the shaded region represents the ladder vertex corrections.
    The dotted lines represent the impurity potential and the crosses represent the impurities.}
    \label{fig:diagram}
\end{figure}

In this section, we demonstrate the calculation of the spin density and spin-current density driven by the differential rotation in linear response to the emergent gauge fields.
The response functions of spin density and spin-current density corresponding to the diagrams shown in Fig.~\ref{fig:diagram}(a) are expressed as
\begin{align}
    K_{\mu \nu}(\omega) = \delta_{\mu 0} \delta_{\nu 0} \frac{\hbar \sigma_c}{2e^2D} + i\omega \frac{ \hbar}{4\pi} \sum_{\bm k} v_{\bm k,\mu} \text{tr} [\sigma^z g^r_+ \sigma^z(v_{\bm k,\nu}+\Lambda_{\nu}^{s})g^a_-],
\end{align}
where the three-point vertices $\Lambda^s_\nu$ are shown in Fig.~\ref{fig:diagram}(b).
Up to the second order in the wavenumber $\bm q$ and frequency $\omega$, the response functions are calculated as
\begin{align}
    K_{00}(\omega)
    &
    = \frac{\hbar \sigma_c}{2e^2D} + i\omega \frac{\hbar}{2\pi} \Bigl[ 1 + \Lambda_0^{s} \Bigr] I_0
    = \frac{\hbar \sigma_c}{2e^2D} \frac{Dq^2 + \tau_s^{-1}}{Dq^2-i\omega + \tau_{s}^{-1}},
    \\
    K_{0j}(\omega) &= i\omega \frac{\hbar}{2\pi} \Bigl[ 1 + \Lambda_0^{s} \Bigr] I_j = -i\omega \frac{\hbar \sigma_c}{2e^2} \frac{iq_j}{Dq^2-i\omega + \tau_{s}^{-1}},
    \\
    K_{i0}(\omega) &= i\omega \frac{\hbar}{2\pi} I_{i} \Bigl[ 1 + \Lambda_0^{s} \Bigr]= -i\omega \frac{\hbar \sigma_c}{2e^2} \frac{iq_i}{Dq^2-i\omega + \tau_{s}^{-1}},
\\
    K_{ij}(\omega) &= i\omega \frac{\hbar}{2\pi} \Bigl[ I_{ij} + I_{i}\Lambda_j^{s} \Bigr]
    = i\omega \frac{\hbar \sigma_{c}}{2e^2} \left( \delta_{ij} - \frac{Dq_iq_j}{Dq^2-i\omega + \tau_{s}^{-1}}  \right),
\end{align}
where the Latin indices represent the spacial directions, i.e., $i=x,y,z$.
Here, the integrations $I_{\mu \nu}$ are defined by
\begin{align}
    I_{\mu \nu} = \sum_{\bm k} v_{\bm k,\mu} v_{\bm k,\nu} g^r_+g^a_-,
\end{align}
and given by
\begin{gather}
    I_0 = I_{00} = \frac{\pi \nu_0}{\hbar \gamma}[1-\tau(Dq^2-i\omega)],\\
    I_i=I_{i0}=I_{0i} = -Diq_i \frac{\pi \nu_0}{\hbar \gamma},\\
    I_{ij} =  \frac{v_F^2}{3} \frac{\pi \nu_0}{\hbar \gamma} 
    \left[
\delta _{ij} + \delta_{ij} \tau \left( i\omega -\frac{3}{5}Dq^2 \right) - \frac{6}{5}\tau D q_i q_j
\right].
\end{gather}
Note that the response functions satisfy the following identities: 
\begin{equation}
    - i\omega K_{00} + iq_j K_{0j} = -i\omega \frac{\hbar \sigma_{c}}{2e^2 D} \frac{\tau_{s}^{-1}}{Dq^2-i\omega + \tau_{s}^{-1}} ,\quad
    - i\omega K_{i0} + iq_j K_{ij} = i\omega \frac{\hbar \sigma_{c}}{2e^2} \frac{\tau_{s}^{-1}}{Dq^2-i\omega + \tau_{s}^{-1}} iq_i .
\end{equation}
The spin density is given by
\begin{align}
    \average{\hat s (\bm q,\omega)} &= K_{00}A_{s,0} + K_{0j} A_{s,j}
\nonumber    \\
    &
    =\frac{\hbar \sigma_{c}}{2e^2 D}
    \frac{\tau_{s}^{-1}}{Dq^2-i\omega + \tau_{s}^{-1}} A_{s,0}
    + \frac{\hbar \sigma_{c}}{2e^2}
    \frac{iq_j}{Dq^2-i\omega + \tau_{s}^{-1}} (-i\omega A_{s,j} - iq_j A_{s,0})
\nonumber    \\
    &= \frac{\hbar \sigma_{c}}{2e^2D}  \frac{\tau_{s}^{-1}}{Dq^2-i\omega + \tau_{s}^{-1}} (-i\omega \Phi).
\end{align}
The spin-current density is given by
\begin{align}
    \average{\hat j_{s,i}(\bm q,\omega)} &= K_{i0}A_{s,0} + K_{ij} 
    A_{s,j}
\nonumber    \\
    &
    = i\omega \frac{\hbar \sigma_{c}}{2e^2}  \frac{\tau_{s}^{-1}}{Dq^2-i\omega + \tau_{s}^{-1}} A_{s,i} \nonumber\\
    &
    \quad + i\omega \frac{\hbar \sigma_{c}}{2e^2} \frac{1}{Dq^2-i\omega + \tau_{s}^{-1}} (-i\omega A_{s,i} - iq_i A_{s,0})
    + i\omega \frac{\hbar \sigma_{c}}{2e^2} \frac{iDq_j}{Dq^2-i\omega + \tau_{s}^{-1}} (iq_i A_{s,j} - iq_j A_{s,i})
\nonumber    \\
    &= i\omega \frac{\hbar \sigma_{c}}{2e^2}  \frac{\tau_{s}^{-1}}{Dq^2-i\omega + \tau_{s}^{-1}} iq_i \Phi.
\end{align}
We note that $(A_{s,0} ,A_{s,i}) = (-i\omega \Phi, iq_i \Phi)$.

	\subsection{Ladder vertex corrections}
	
	In this section, we calculate ladder vertex corrections due to the impurity scattering and spin-orbit scattering.
	First, we define the elementary vertex $f_{ab}$ shown in Fig.~\ref{fig:vertex}(a) corresponding to the coupling to the single impurity:
	\begin{align}
	    f_{ab} = \delta_{ab} + i\hbar \lambda_{\text{so}} (\bm k\times \bm k') \cdot \bm \sigma_{ab},
	\end{align}
	where the Latin indices $a,b$ describe the spin space.
	The proper four-point vertex $\Gamma^0$ shown in the first term on the right-hand side of Fig.~\ref{fig:vertex}(b) is calculated as
	\begin{align}
	    \Gamma^0_{ab,cd}=n_iu_i^2 \average{f_{ad}f_{cb}}_{\text{FS}} = \frac{\hbar}{\pi \nu_0} \left( \gamma_0 \delta_{ad}\delta_{cb} + \frac{1}{3}\gamma_{\text{so}} \bm \sigma_{ad}\cdot \bm \sigma_{cb} \right),
	\end{align}
	where $\average{\cdots}_{\text{FS}}$ means averaging over at the Fermi surface, $\hbar \gamma_0=\pi n_iu_i^2\nu_0$ is the damping due to the impurity scattering, and $\gamma_{\text{so}}= \hbar^2 \lambda_{\text{so}}^2 k_F^4 \gamma_0$ is the damping due to the spin-orbit scattering.
	The four-point vertex shown in Fig.~\ref{fig:vertex}(b) is determined by the following Dyson equation:
	\begin{align}
	    \Gamma_{ab,cd}(\bm q) &= \Gamma^0_{ab,cd} + \Gamma^0_{ab,ef} I_0(\bm q) \Gamma_{fe,cd}(\bm q) \nonumber 
     \\
     &= \Gamma _{c}(\bm q) \delta _{ab} \delta _{cd} + \Gamma _{s}(\bm q) \bm \sigma _{ab} \cdot \bm \sigma _{cd},
	\end{align}
where $a, \ldots ,f$ are the spin indices, and
\begin{align}
\Gamma _{c}(\bm q) &= \frac{\hbar}{4\pi \nu_0 \tau ^2} \frac{1}{Dq^2-i\omega }, \\
\Gamma _{s}(\bm q) &= \frac{\hbar}{4\pi \nu_0 \tau ^2} \frac{1-\frac{\tau}{\tau _{s}}}{Dq^2-i\omega  + \tau _{s}^{-1}}.
\end{align}
Therefore, the three-point vertices $\Lambda ^{s}_{\nu}$ are calculated by
\begin{align}
\sigma ^{\alpha}_{ab}\Lambda ^{s}_{\nu }(\bm q) &=\sigma ^{\alpha}_{dc} \Gamma _{ab, cd}(\bm q) I_{\nu},
\end{align}
and given by
\begin{align}
\Lambda ^{s}_{0}(\bm q) &= 
\dfrac{1}{\tau (Dq^2-i\omega + \tau^{-1}_{s}) } -1,
\\
\Lambda ^{s}_j(\bm q) &= -\dfrac{Diq_j}{\tau (Dq^2-i\omega + \tau^{-1}_{s}) }.
\end{align}

\begin{figure}
    \centering
    \includegraphics[width=120mm]{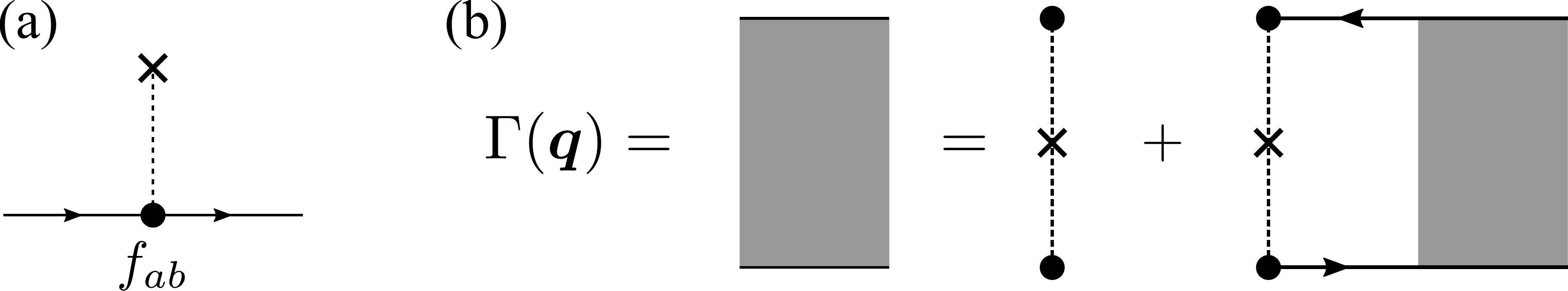}
    \caption{(a) The elementary vertex due to the single impurity. (b) The four-point vertex in the ladder approximation.}
    \label{fig:vertex}
\end{figure}

\red{
\section{Derivation of Navier-Stokes equation for two-dimensional cylindrical flow}
In this section, we derive the Navier-Stokes equation for two-dimensional steady flow with concentric circular streamlines and demonstrate that, in this case, the Navier-Stokes equations are explicitly independent of viscosity.
The Navier-Stokes equation in the cylindrical coordinate representation is given by
\begin{align}
    \pdv{v_r}{t} + v_r \pdv{v_r}{r} + \frac{v_\varphi}{r} \pdv{v_r}{\varphi} + v_z \pdv{v_r}{z} - \frac{v_\varphi^2}{r} &= -\frac{1}{\rho} \pdv{p}{r} + \nu \qty( \Delta v_r - \frac{v_r}{r^2} - \frac{2}{r^2} \pdv{v_\varphi}{\varphi} ),
\\
    \pdv{v_{\varphi}}{t} + v_r \pdv{v_{\varphi}}{r} + \frac{v_\varphi}{r} \pdv{v_\varphi}{\varphi} + v_z \pdv{v_\varphi}{z} + \frac{v_r v_\varphi}{r} &= -\frac{1}{\rho r} \pdv{p}{\varphi} + \nu \qty( \Delta v_\varphi - \frac{v_\varphi}{r^2} + \frac{2}{r^2} \pdv{v_r}{\varphi} ),
\\
    \pdv{v_z}{t} + v_r \pdv{v_z}{r} + \frac{v_\varphi}{r} \pdv{v_z}{\varphi} + v_z \pdv{v_z}{z}  &= -\frac{1}{\rho} \pdv{p}{z} + \nu \Delta v_z,
\end{align}
where $\Delta$ represents the Laplacian, given by
\begin{align}
    \Delta = \pdv[2]{r} + \frac{1}{r}\pdv{r} + \frac{1}{r^2}\pdv[2]{\varphi} + \pdv[2]{z}.
\end{align}
For the two-dimensional flow with concentric circular streamlines, the velocity vector can be represented as $\bm v=(0,v_\varphi,0)$, and the derivatives with respect to $\varphi$ and $z$ vanish, i.e., $\partial/\partial\varphi = \partial/\partial z=0$.
Then, the Navier-Stokes equation turn to be represented as
\begin{align}
    \pdv{v_{\varphi}}{t}= \nu \qty( \pdv[2]{v_\varphi}{r} + \frac{1}{r}\pdv{v_\varphi}{r} - \frac{v_\varphi}{r^2} ).
\end{align}
In the case of steady flow, i.e., $\partial/\partial t=0$, the Navier-Stokes equation come to be
\begin{align}
    \pdv[2]{v_\varphi}{r} + \frac{1}{r}\pdv{v_\varphi}{r} - \frac{v_\varphi}{r^2} =0.
\end{align}
As can be seen from the above analysis, the specificity of the two-dimensional steady flow with concentric streamlines renders the Navier-Stokes equations explicitly independent of viscosity.
}

\section{Distinction between our spin current generation mechanism and conventional mechanisms}
%\red{
In this section, we note the differences between previous studies and the present results on spin current generation using torsional oscillation. 

Previous studies, such as references (a) M. Matsuo et al., Phys. Rev. B87, 180402 (2013) and (b) Phys. Rev. B96, 020401 (2017), have proposed the concept based on spin-vorticity coupling. According to these theories, when calculating spin current generation from torsional oscillation, we can clearly distinguish our current results by the frequency dependence and the presence or absence of phenomenological parameters, as summariezed below.

\begin{enumerate}
    \item \textbf{Frequency Dependence}: \\
    The results of (a) can be distinguished from our current results by frequency dependence. In the case of (a), the amplitude of the spin current is proportional to the fourth power of the frequency, whereas our current results show that it is proportional to the third power of the frequency.

    \item \textbf{Dependence on Phenomenological Parameter}: \\
    The major difference between the results of (b) and our current results is that (b) includes a phenomenological parameter that originates from a fluctuation of vorticity, whereas our current results do not require phenomenology. 
    Therefore, our work is the first result that allows a quantitative comparison
between experiment and theory.
%Therefore, for the first time, our results allow a precise comparison between theory and experiment. In other words, the decisive difference is that our results do not depend on phenomenology.
\end{enumerate}

These differences arise from the variations in the source term of the spin diffusion equation derived from the theory. Specifically, 
(a) states that the source term is proportional to the derivative of vorticity, $\partial_t \tilde{\omega}$, where $\tilde{\omega}$ is vorticity. 
In contrast, (b) indicates that the source term is proportional to the vorticity and includes a phenomenological parameter as $\zeta \tilde{\omega}$, where $\zeta$ is the phenomenological parameter originating from a fluctuation of vorticity. 
Our current results, on the other hand, demonstrate that the source term is phenomenology-free and proportional to the angular velocity of the differential rotation: 
\begin{equation}
    \frac{\partial s}{\partial t}+{\bm \nabla} \cdot {\bm j}_s + \frac{s}{\tau_{s}} = \frac{\hbar \red{\sigma_0}}{2e^{2}D \tau_s}\partial_t\Phi ,
\label{eq:spin_cons_viloation}
\end{equation}
where $s$ is the spin density, $\boldsymbol{j}_s$ is the spin current density, $\tau_s$ is the spin relaxation time, $\sigma_0$ is the Drude conductivity,  $D$ is the diffusion constant, and $\partial_t \Phi$ is the angular velocity of the differential rotation. 
As a result, the spin current is given by
        \begin{align}
            \bm j_s(\bm x,t) = -\frac{\hbar \red{\sigma_0}}{2e^2} \bm \nabla \partial_t \Phi (\bm x,t).
            \label{js}
        \end{align}
%Since $\sigma_0$ depends on the temperature, we expect that our spin current has the same temperature dependence. Measuring spin current at various temperatures may give us information on the origin of the spin current. 
From this equation, we can see that our spin current depends only on $\hbar$, $e$, and the Drude conductivity but does not involve any phenomenological parameters.

\end{widetext}

\end{document}